\documentstyle[12pt]{article}
\newtheorem{theorem}{Theorem}
\begin{document}
\title{
 General-Relativistic Thomas-Fermi model
 }
\author{
Neven Bili\'c$^1$
and
Raoul D.~Viollier$^2$
 \\
 $^1$Rudjer Bo\v{s}kovi\'{c} Institute, 10000 Zagreb, Croatia \\
E-mail: bilic@thphys.irb.hr \\
$^2$Department of Physics,
University of Cape Town, \\ Rondebosch 7701, South Africa,
E-mail: viollier@physci.uct.ac.za
}
\maketitle
\date{\today}
\begin{abstract}
 A system of self-gravitating
 massive fermions
 is studied
in the framework
of the general-relativistic Thomas-Fermi model.
We study the properties of the free energy functional
and its relation to
Einstein's field equations.
A self-gravitating fermion gas
we then describe by
a set of Thomas-Fermi type  self-consistency
equations.
\end{abstract}
%

%

 Thermodynamical properties of the self-gravitating
 fermion gas have been extensively studied
 in the framework of the Thomas-Fermi model
[1-6].
 The system was investigated in the nonrelativistic
 Newtonian limit.
   The canonical and grand-canonical ensembles
     for such a system
  have been shown to
have a nontrivial thermodynamical limit~\cite{thi,her1}.
Under certain conditions this system
 will undergo a phase transition
 that is accompanied by a gravitational collapse~\cite{her1,mes}
 which may have important astrophysical and cosmological
 implications~\cite{bil1,bil2}.

 In this paper we formulate the general-relativistic
 version of the model.
 The effects of general relativity become important
 if the total rest-mass of the system is close to the
 Oppenheimer-Volkoff limit \cite{opp}.
 There are three main features that distinguish the relativistic
 Thomas-Fermi theory
  from the Newtonian one:
  {\it i}) the equation of state is relativistic
  {\it ii}) the temperature and chemical potential are
  metric dependent local quantities
  {\it iii}) the gravitational potential satisfies Einstein's field
  equations (instead of Poisson's equation).

Let us first discuss the general properties
of a canonical, self-gravitating relativistic fluid.
Consider a nonrotating fluid consisting of $N$ particles
in
a spherical volume of radius $R$ in equilibrium
at non-zero temperature.
We denote by
 $u_{\mu}$ ,
$p$, $\rho$, $n$ and $\sigma$ the velocity, pressure,
energy density, particle number density and
entropy density of the fluid.
A canonical ensemble is subject to the constraint
that the number of particles
\begin{equation}
\int_{\Sigma} n\, u^{\mu}d\Sigma_{\mu}
=N
\label{eq26}
\end{equation}
should be fixed.
The spacelike hypersurface
$\Sigma$ that contains
the fluid is orthogonal to the time-translation
Killing vector field $k^{\mu}$
which is related to the velocity of the fluid
\begin{equation}
k^{\mu}=\xi u^{\mu}\, ;  \;\;\;\;\;\;
\xi=(k^{\mu}k_{\mu})^{1/2}.
\label{eq50}
\end{equation}
The metric generated by the mass distribution
is  static, spherically symmetric and asymptotically
flat, i.e.
\begin{equation}
ds^2=\xi^2 dt^2 -\lambda^2 dr^2 -
     r^2(d\theta^2+\sin \theta d\phi^2).
\label{eq00}
\end{equation}
$\xi$ and $\lambda$ may be represented in terms of the
gravitational potential and mass
\begin{equation}
\xi=e^{\varphi (r)},
\label{eq01}
\end{equation}
\begin{equation}
\lambda=\left(1-\frac{2{\cal{M}}(r)}{r}\right)^{-1/2}
\label{eq10}
\end{equation}
with
\begin{equation}
{\cal{M}}(r)=\int^r_0 dr'\, 4\pi r'^2 \rho(r') \, .
\label{eq11}
\end{equation}

 The temperature $T$ and chemical potential $\mu$ are
  metric dependent local quantities.
  Their space-time dependence may be derived from
the equation of hydrostatic equilibrium
\cite{lan}
\begin{equation}
\partial_{\nu}p=-(p+\rho)\xi^{-1}\partial_{\nu}\xi ,
\label{eq17}
\end{equation}
and the thermodynamic identity (Gibbs-Duhem relation)
\begin{equation}
 d\frac{p}{T}=
 n d\frac{\mu}{T}-\rho d\frac{1}{T}.
\label{eq18}
\end{equation}
The condition that the heat flow and diffusion vanish
\cite{isr}
\begin{equation}
  \frac{\mu}{T}={\rm const}
\label{eq19}
\end{equation}
together with (\ref{eq17}) and (\ref{eq18}) implies
\begin{equation}
T \xi=T_0\, ; \;\;\;\;\;\;
\mu \xi=\mu_0  \, ,
\label{eq21}
\end{equation}
where $T_0$ and $\mu_0$ are constants equal to the
temperature and chemical potential at infinity.
The temperature $T_0$ may be chosen arbitrarily as the temperature
of the heat-bath.
The quantity
$\mu_0$
in a canonical ensemble is
an implicit functional of $\xi$ owing to
the constraint (\ref{eq26}).
First equation in (\ref{eq21}) is the well known Tolman condition
for thermal equilibrium in a gravitational
field \cite{tol}.

Following Gibbons and Hawking \cite{gib} we postulate
the free energy of the canonical ensemble as
\begin{equation}
F=M-\int_{\Sigma} T\sigma \, k^{\mu}d\Sigma_{\mu} \, ,
\label{eq30}
\end{equation}
where $M$ is the total mass as measured from infinity.
 The entropy density of a relativistic fluid may be expressed as
\begin{equation}
\sigma=\frac{1}{T}(p+\rho-\mu n).
\label{eq16}
\end{equation}
Based on equation (\ref{eq21}) the free energy may be written
in the form analogous to ordinary thermodynamics
\begin{equation}
F=M-T_0 S
\label{eq60}
\end{equation}
with  $M={\cal{M}}(R)$
and the total entropy $S$ defined as
\begin{equation}
S = \int_0^R dr\,4\pi r^2 \lambda
\frac{1}{T}(p+\rho)-\frac{\mu_0}{T_0} N  ,
\label{eq70}
\end{equation}
where we have employed the spherical symmetry to
replace the proper volume integral as
\begin{equation}
\int_{\Sigma} u^{\mu}d\Sigma_{\mu}
= \int_0^R dr 4\pi r^2 \lambda .
\label{eq80}
\end{equation}

The following theorem demonstrates how the
extrema of the free energy are related to
the solutions of
 Einstein's field equation.

\begin{theorem}
Among all momentarily
static, spherically symmetric configurations
$\{\xi(r),{\cal{M}}(r)\}$
which for a given temperature $T_0$ at infinity
 contain a specified number of particles
\begin{equation}
 \int_0^R 4\pi r^2 dr \, \lambda(r)  n(r) = N
\label{eq25}
\end{equation}
within a spherical volume of a given radius
 $R$,
those and only those  configurations
that
extremize the quantity F defined by
{\rm (\ref{eq60})}
 will
 satisfy Einstein's field equation
\begin{equation}
\label{eq22}
\frac{d\xi}{dr}=\xi\frac{{\cal{M}}+4\pi r^3 p}{r(r-2{\cal{M}})} \, ,
\end{equation}
with the boundary condition
\begin{equation}
\xi(R)=\left(1-\frac{2 M}{R}\right)^{1/2}.
\label{eq23}
\end{equation}
\end{theorem}
{\bf Proof.}
By making use of the identity (\ref{eq18}),
 and the fact
that $\delta(\mu/T)=\delta(\mu_0/T_0)$
and that $N$ is fixed by the constraint (\ref{eq25}),
from equations (\ref{eq60}) and (\ref{eq70})
we find
\begin{equation}
\delta F= \delta M -
\int_0^R dr\, 4\pi r^2 \frac{T_0}{T}(p+\rho)
\delta \lambda
-  \int_0^R dr\, 4\pi r^2 \lambda \frac{T_0}{T} \delta\rho \, .
\label{eq90}
\end{equation}
The variations $\delta\lambda$ and $\delta\rho$
 can be expressed in terms of the variation
$\delta {\cal{M}}(r)$
and its derivative
\begin{equation}
\frac{d\delta {\cal{M}}}{dr} =4\pi r^2 \delta\rho.
\label{eq93}
\end{equation}
This  gives
\begin{equation}
\delta F= \delta M -
\int_0^R dr\, 4\pi r^2
 \frac{T_0}{T}(p+\rho)
\frac{\partial\lambda}{\partial {\cal{M}}}
\delta {\cal{M}}
-\int_0^R dr\, \lambda\frac{T_0}{T}\frac{d\delta {\cal{M}}}{dr}.
\label{eq91}
\end{equation}
By partial integration of the last term
and replacing $T_0/T$ by $\xi$, we find
\begin{equation}
\delta F =
\left[1-\lambda(R)\xi(R)\right]\delta M
- \int_0^R dr\, \left[4\pi r^2 \xi (p+\rho)
\frac{\partial\lambda}{\partial {\cal{M}}}
-\frac{d}{dr}(\lambda\xi)\right]\delta {\cal{M}} \, ,
\label{eq94}
\end{equation}
where $\delta {\cal{M}}(r)$ is an arbitrary variation
on the interval $[0,R]$,
except for
the constraint
$\delta {\cal{M}}(0)=0$.
Therefore $\delta F$ will vanish if and only if
\begin{equation}
4\pi r^2 \xi (p+\rho)
\frac{\partial\lambda}{\partial {\cal{M}}}
-\frac{d}{dr}(\lambda\xi) =0
\label{eq95}
\end{equation}
and
\begin{equation}
1-\lambda(R)\xi(R) =0.
\label{eq96}
\end{equation}
Using (\ref{eq10}) and (\ref{eq11}), we can write
equation (\ref{eq95}) in the form (\ref{eq22}),
and equation (\ref{eq96}) gives the desired boundary condition
 (\ref{eq23}).
Thus,
$\delta F=0$ if and only if a configuration
$\{\xi,{\cal{M}}\}$ satisfies equation (\ref{eq22})
with (\ref{eq23})
as was to be shown.
\\
{\it Remark 1.}
A solutions to equation (\ref{eq22})
is dynamically stable if
the free energy assumes a minimum.
\\
{\it Remark 2.}
Our Theorem 1 is a finite temperature generalization
of the result obtained for
cold, catalyzed matter \cite{har}.

We now proceed to the formulation of the general-relativistic
Thomas-Fermi model.
Consider the case of a self-gravitating
gas consisting of $N$ fermions with the mass
$m$ contained in a sphere of radius $R$.
The equation of state may be represented
in a parametric form using
the well known momentum integrals over
 the Fermi distribution function
\cite{ehl}
\begin{equation}
n   = g \int^{\infty}_{0} \frac{d^3q}{(2\pi)^3}\,
\frac{1}{1+e^{E/T-\mu/T}} \, ,
\label{eq13}
\end{equation}
\begin{equation}
\rho = g \int^{\infty}_{0} \frac{d^3q}{(2\pi)^3}\,
\frac{E}{1+e^{E/T-\mu/T}} \, ,
\label{eq14}
\end{equation}
\begin{equation}
p = g T \int^{\infty}_{0} \frac{d^3q}{(2\pi)^3}\,
\ln (1+e^{-E/T+\mu/T}) \, ,
\label{eq15}
\end{equation}
where
$g$ denotes the spin degeneracy factor,
$T$ and $\mu$ are local temperature and chemical potential,
respectively,
as defined in equation (\ref{eq21}),
 and $E=\sqrt{m^2+q^2}$.
 Introducing a single parameter
\begin{equation}
\alpha=
\frac{\mu}{T}
=\frac{\mu_0}{T_0} \, ,
\label{eq100}
\end{equation}
and the substitution
\begin{equation}
\xi=
\frac{\mu_0}{m}\psi \, ,
\label{eq102}
\end{equation}
 equations (\ref{eq13})-(\ref{eq15})
 may be written in the form
\begin{equation}
n   = g \int^{\infty}_{0} \frac{d^3q}{(2\pi)^3}\,
\frac{1}{1+e^{(E\psi-m)\alpha}} \, ,
\label{eq104}
\end{equation}
\begin{equation}
\rho= g \int^{\infty}_{0} \frac{d^3q}{(2\pi)^3}\,
\frac{E}{1+e^{(E\psi-m)\alpha}} \, ,
\label{eq106}
\end{equation}
\begin{equation}
p   = g \int^{\infty}_{0} \frac{d^3q}{(2\pi)^3}\,
\frac{q^2}{3E}\frac{1}{1+e^{(E\psi-m)\alpha}} \, ,
\label{eq108}
\end{equation}
 Field equations are given by
\begin{equation}
\frac{d\psi}{dr}=\psi\frac{{\cal{M}}+4\pi r^3 p}{r(r-2{\cal{M}})} \, ,
\label{eq42}
\end{equation}
\begin{equation}
\frac{d{\cal{M}}}{dr}=4\pi r^2 \rho,
\label{eq43}
\end{equation}
with the boundary conditions
\begin{equation}
\psi(R)=\frac{m}{\mu_0}\left(1-\frac{2 {\cal{M}}(R)}{R}\right)^{1/2}
\, ; \;\;\;\;\;
{\cal{M}}(0)=0.
\label{eq44}
\end{equation}
Finally, the constraint (\ref{eq26}) may be written as
\begin{equation}
\int_0^Rdr\, 4\pi r^2 (1-2{\cal{M}}/r)^{-1/2}\, n(r)=N .
\label{eq45}
\end{equation}
 Given the ratio $\alpha$, the radius $R$, and
 the number of fermions $N$,
the set of self-consistency
equations (\ref{eq104})-(\ref{eq45}) defines
the Thomas-Fermi equation.
One additional important requirement is that a solution
of the self-consistency equations (\ref{eq104})-(\ref{eq45})
should minimize
the free energy
defined by (\ref{eq60}).

We now show that
a solution of
the Thomas-Fermi
equation exists
provided the number of fermions
is smaller than a
certain number $N_{\rm max}$
that depends on
$\alpha$ and $R$.
From  (\ref{eq106}) and
  (\ref{eq108})   it follows that
for any $\alpha>0$,
the equation of state $\rho(p)$ is an
infinitely smooth function and
$d\rho /dp > 0$ for $p > 0$.
Then, as shown by
Rendall and Schmidt
\cite{ren},
 there exist for any value of
the central density $\rho_0$
a unique static, spherically symmetric solution of
field equations
with $\rho \rightarrow 0 $  as
$r$ tends to infinity.
In that limit
${\cal M}(r)\rightarrow\infty$,
as may easily be seen
by analysing
the
$r\rightarrow \infty$
limit of
equations (\ref{eq42}) and (\ref{eq43}).
However, the enclosed mass
$M$ and the number of fermions $N$
within a given radius $R$ will be finite.
We can then cut off the matter
from $R$ to infinity
and join on
the empty space Schwarzschild
solution
by making use of
equation
 (\ref{eq44}).
 This equation together with
 (\ref{eq100})
 fixes the chemical potential
 and the temperature at infinity.
Furthermore, it may be shown that
our equation of state obeys
a $\gamma$-low
asymptotically  at high densities,
i.e.,
$\rho=$ const $n^{\gamma}$
and $p=(\gamma-1) \rho$,
with
$\gamma=4/3$.
It is well known
\cite{har} that
in this case, there exist a limiting configuration
$\{ \psi_{\infty}(r),{\cal{M}}(r)_{\infty}\}$
such that $M$ and $N$
approach non-zero values $M_{\infty}$
and $N_{\infty}$,
respectively,
as the central density
$\rho_{0}$
tends to infinity.
Thus, the quantity
 $N$ is a continuous function of
 $\rho_{0}$
on the interval $0 \leq \rho_0 < \infty$,
with $N=0$ for
 $\rho_{0}=0$,  and
 $N=N_{\infty}$
 as
 $\rho_{0}\rightarrow\infty$.
The range of $N$
depends on
$\alpha$ and $R$
and its
upper bound
 may be denoted by
$N_{\rm max}(R,\alpha)$.
Thus, for given
$\alpha$, $R$
and
$N<N_{\rm max}(R,\alpha)$
the set of self-consistency
equations (\ref{eq104})-(\ref{eq45}) has
at least one solution.

Next we  show  that, in the
Newtonian limit, we recover the
 nonrelativistic Thomas-Fermi equation.
Using
the nonrelativistic chemical potential
$\mu_{NR}=\mu_0-m$
and
 the approximation
$\xi=e^{\varphi}\simeq 1+\varphi$,
$E\simeq m+q^2/2m$
and
  ${\cal{M}}/r \ll 1$ ,
 we find the usual Thomas-Fermi self-consistency
 equations \cite{mes,bil1}
\begin{equation}
n=\frac{\rho}{m}
 = g \int^{\infty}_{0} \frac{d^3q}{(2\pi)^3}\,
\left(1+\exp(\frac{q^2}{2mT_0}+\frac{m}{T_0}\varphi
 -\frac{\mu_{NR}}{T_0}) \right)^{-1} \, ,
\label{eq49}
\end{equation}
\begin{equation}
\frac{d\varphi}{dr}=\frac{{\cal{M}}}{r^2} \, ;
\;\;\;\;
\frac{d{\cal{M}}}{dr}=4\pi r^2 \rho \, ,
\label{eq41}
\end{equation}
\begin{equation}
\varphi(R)=-\frac{m N}{R}
\, ; \;\;\;
{\cal{M}}(0)=0,
\label{eq47}
\end{equation}
\begin{equation}
\int_0^R dr\,4\pi r^2 n(r)=N.
\label{eq46}
\end{equation}
The free energy  (\ref{eq60}) in the Newtonian limit yields
\begin{equation}
F=m N +\mu_{NR} N - \frac{1}{2}\int_0^R dr \, 4\pi r^2 n\varphi
-\int_0^R dr \, 4\pi r^2 p
\label{eq40}
\end{equation}
with
\begin{equation}
 p= g T_0\int^{\infty}_{0} \frac{d^3q}{(2\pi)^3}\,
\ln\left(1+\exp(-\frac{q^2}{2mT_0}-\frac{m}{T_0}\varphi
 +\frac{\mu_{NR}}{T_0}) \right) \, ,
\label{eq48}
\end{equation}
which, up to a constant, equals the Thomas-Fermi free
energy \cite{her2}.

A straightforward thermodynamic limit
$N\rightarrow\infty$
as discussed by
Hertel, Thirring and Narnhofer \cite{her1,her2}
is in our case not directly applicable.
First, in contrast to the non-relativistic case,
 there exists, as we have demonstrated, a limiting configuration
with maximal $M$ and $N$.
Second,
the scaling properties of the relativistic
Thomas-Fermi equation are quite distinct from
the nonrelativistic one.
The following scaling property can be easily shown:
  If the configuration
 $\{\psi(r),{\cal{M}}(r)\}$ is a solution of the self consistency
 equations (\ref{eq104})-(\ref{eq45}), then the
 configuration
 $\{\tilde{\psi}=\psi(A^{-1}r),\tilde{{\cal{M}}}
 =A{\cal{M}}(A^{-1}r);A>0\}$
 is also a solution with
 the rescaled
 fermion number
 $\tilde{N}=A^{3/2}N$,
 radius
 $\tilde{R}=AR$,
 asymptotic temperature
 $\tilde{T_0}=A^{-1/2}T_0$,
 and fermion mass
 $\tilde{m}=A^{-1/2}m$.
 The free energy is then rescaled as
 $\tilde{F}=AF$.
 Therefore, there exist a thermodynamic limit
 of
 $N^{-2/3}F$,
 with
 $N^{-2/3}R$,
 $N^{1/3}T_0$,
 $N^{1/3}m$
 approaching constant values
  when
 $N\rightarrow\infty$.
 In that limit
 the Thomas-Fermi equation becomes exact.

 It is obvious that application of this model
 to astrophysical systems
 should work very well if the interactions
 among individual particles are negligible.
This applies, for example, to
weakly interacting quasidegenerate heavy neutrino or
neutralino matter [6,7,16-19].
or perhaps even to collisionless stellar systems
\cite{shu,chav}.
\subsection*{Acknowledgment}
We acknowledge useful discussions with
 D. Tsiklauri.
 This work  was supported by
 the Foundation for Fundamental Research
 (FFR) and the Ministry of Science and Technology of the
 Republic of Croatia under Contract
 No. 00980102.
\vspace{0.2in}
%
%

%


\begin{thebibliography}{99}
%
\bibitem{thi}
W. Thirring,  Z. Phys. {\bf 235} (1970) 339.
\bibitem{her1}
P. Hertel and W. Thirring, Comm. Math. Phys. {\bf 24} (1971) 22;
P. Hertel and W. Thirring,
``Thermodynamic Instability of a System of
Gravitating Fermions", in
{\it Quanten und Felder}, edited by H. P. D\"urr
(Vieweg, Braunschweig, 1971).
\bibitem{her2}
P. Hertel, H. Narnhofer and
W. Thirring, Comm. Math. Phys. {\bf 28} (1972) 159.
\bibitem{bau}
B. Baumgartner,
 Comm. Math. Phys. {\bf 48} (1976) 207.
\bibitem{mes}
J. Messer, J. Math. Phys. {\bf 22} (1981) 2910.
\bibitem{bil1}
N. Bili\'c and R.D. Viollier,
Phys. Lett. {\bf B 408}
(1997) 75;
N. Bili\'c and R.D. Viollier,
Nucl. Phys. {\bf B} (Proc. Suppl.) {\bf 66}
(1998) 256.
%
\bibitem{bil2}
N. Bili\'c, D. Tsiklauri,
and R.D. Viollier,
Prog. Part. Nucl. Phys. {\bf 40}
(1998) 17.
%
\bibitem{opp}
J.R. Oppenheimer and G.M. Volkoff,
Phys. Rev.
{\bf 55} (1939) 374.
%
\bibitem{lan}
L.D. Landau, E.M. Lifshitz,
{\em Fluid Mechanics},
(Pergamon, Oxford, 1959) p. 503.
\bibitem{isr}
W. Israel,
Ann. Phys. {\bf 100}
(1976) 310
\bibitem{tol}
R.C. Tolman,
{\em Relativity Thermodynamics and Cosmology},
(Clarendon, Oxford, 1934) p. 312-317.
\bibitem{gib}
G.W. Gibbons and S.W. Hawking,
Phys. Rev.
{\bf D55} (1977) 2752.
%
\bibitem{har}
B.K. Harrison, K.S. Thorne, M. Wakano
and J.A. Wheeler,
{\em Gravitation Theory and Gravitational Collapse},
(The University of Chicago Press, Chicago, 1965).
ch. 3-5.
%
\bibitem{ehl}
J. Ehlers,
Survey of General Relativity Theory,
in {\em Relativity, Astrophysics and Cosmology},
ed W. Israel
(D. Reidel Publishing Company,
Dordrecht/Boston, 1973), sect. 3.
%
\bibitem{ren}
A.D. Rendall and B.G. Schmidt,
Class. Quantum Grav. {\bf 8}
(1991) 985
%
\bibitem{cha}
W.Y. Chau, K. Lake,
and J. Stone, Ap. J. {\bf 281} (1984) 560
%
\bibitem{kul}
A. Kull, R.A. Treumann, and H. B\"ohringer
Ap. J. {\bf 466} (1996) L1.
%
\bibitem{tsi}
D. Tsiklauri
and R.D. Viollier,
Ap. J. {\bf 500} (1998) 591.
%
\bibitem{bil3}
N. Bili\'c, F. Munyaneza, and R.D. Viollier,
astro-ph/9801262,
Phys. Rev. {\bf D59} (1999) 024003.
%
\bibitem{shu}
F.H Shu,
Ap. J. {\bf 225} (1978) 83.
%
\bibitem{chav}
P.-H. Chavanis and J. Sommeria,
MNRAS {\bf 296} (1998) 569.
%
%
\end{thebibliography}
\end{document}